# The Influence of Epistemic Communities on International Political Negotiations about the Space Debris Problem

## Miloslav Machoň

**Abstract:** Since the 1970's the debate about the rising importance of transnational relations has existed in international relations. Apart from states, related research also focuses on other actors, including epistemic communities. The article uses the concept of epistemic communities and finds whether the activity of epistemic communities determines the process of the international management of outer space in the case of the political negotiations relating to space debris in UNCOPUOS and UNOOSA. The activity of epistemic communities exists in the political negotiations relating to space debris in UNCOPUOS and UNOOSA, but it has not been reflected in the related scholarly literature. Epistemic communities from the non-governmental organizations IAF, COSPAR and IISL contributed to setting the space debris problem on the agenda of UNCOPUOS. Also, under the influence of epistemic communities from the governmental organization IADC, UNCOPUOS adopted guidelines preventing the creation of further amounts of space debris.
**Key words:** epistemic communities, space debris, UNOOSA, UNCOPUOS, IAF, COSPAR, IADC, IISL.

Since the late fifties of the last century, the debate on the gradual transformation of international politics (Wolfers 1959; Burton 1962) has been rising in the international relations theory. This debate intensified in connection with the onset of the theory of complex interdependence (Nye-Keohane 1971), which formed the basis for a later thesis on weakening the importance of the state in international relations (Strange 1996). It has been shown that the internal and external sovereignty of the state has been gradually disrupted by various external factors such as the impact of international financial markets (Gilpin 2011) or by extending the various rules and standards adopted by international organizations and international conferences (Karns - Mingst 2004). At the same time, the study of actors of transnational relations, i.e. all such contacts and interactions that cross the national borders and are not controlled by governmental foreign policy bodies (Keohane-Nye 1971: 330-331), was also rapidly developing. In addition to the state, as a traditional actor of international relations, new types of actors, including multinational corporations (Strange 1996, Rugman 2008), transnational NGOs (Clarke 1998, Karns - Mingst 2004), transnational advocacy networks (Finnemore - Sikkink 1998) civil society (Price 1998, Kaldor 2003).

In the context of growing interest in non-state actors, political science has focused on study of networks of professionals with recognized expertise and professional qualifications in a particular field of science (Fischer 1990, White 1994, Jasanoff 2003, Bucchi-Neresini 2008, Irwin 2008). The importance of networks of professionals in international politics was elaborated by Haas (1990; 1992), which collectively referred to as "epistemic communities" (Haas 1990: 2). He was inspired by Kuhn (1962) who explored the links between individuals within the scientific community [1] and found them in a shared paradigm or a set of shared beliefs and procedures for conducting scientific research. At the same time, he drew from the research of the relationship between knowledge and power, which was realized by Foucault (1970, 1980) already in the 1970s. Significant development of Foucault's original relationship between knowledge and power [2] and Kuhn's concept of the scientific community was Ruggie's work on the circumstances of the emergence of epistemic communities. Epistemic communities may, according to Ruggie, arise not only on the basis of shared paradigms and procedures of scientific research but also from a power position or as a consequence of a revaluation of the meaning of technology (Ruggie 1975: 567-570). The significance of Haas's epistemic communities in international politics has been examined in a study on international regimes contributing to the environmental protection of the Mediterranean (Haas, 1990) and in an introduction to the International Organization in more detail (Haas 1992: 1-35). Haas and Adler also have shown that epistemic communities are creating knowledge that is important for policymaking, especially in conditions of uncertainty.[3] The importance of epistemic communities in international politics is based on the fact that they bring new ideas into international politics and spread them (Haas 1992; Adler - Haas 1992: 372-378).

Existing studies that use the Adler and Haas concepts of epistemic communities focus on the importance of the concept of international relations theory (Hynek 2004) or deal with sub-areas of environmental policy (Toke 1999, Jehlička 2000, Gough-Shackley 2001), public and economic policies (Waarden - Drahos 2002; Merkle 2013) and the process of European integration (Radaelli 1999; Zito 2001). Many spheres of international politics remain uncovered where decisions need to be made in conditions of uncertainty or where decision-making requires highly professional knowledge.

The present text attempts to fill this gap in the knowledge of the importance of epistemic communities, at least in part, by analyzing the issue of space. The text asks the following: Do professional organizations, as an epistemic community, influence international cooperation in space management? Although the theme of space junk may seem marginal in international relations research, it must be emphasized that it has very important security and economic consequences. The attention of the Czech and Slovak professional audiences also deserves this topic because the activities of Czechoslovak and Czech experts contributed significantly to the problem of space debris in UNOOSA and UNCOPUOS (cf. Perek 2002).

Due to the scientific and technological demands of research and the use of space, it can be assumed that the work of epistemic communities is contributing to the development of outer space management, but only so far none of the studies have been dealt with. The text therefore tests the validity of the hypothesis that the activity of epistemic communities affects international co-operation in space. It uses an example of the political negotiations on space shattering at the United Nations Office for Outer Space Affairs (UNOOSA) and the United Nations Committee on the Peaceful Uses of Outer Space (UNCOPUOS).

The text is divided into three chapters. The first chapter summarizes and develops the theoretical framework for recognizing epistemic communities and defines the theme of the importance of epistemic communities in international politics (Haas 1992; Adler - Haas 1992; Cross 2013). The second chapter analyses the active actors in the UN political negotiations on space debris as an epistemic community. The third chapter discusses the influence of epistemic communities on the discourse of political negotiations from 1978 to the present. The analysis of active actors and their influence on the course of political negotiations is based mainly on the constituent documents of the individual organizations, as well as on the records of the political negotiations and the expert studies presented at the UNOOSA and UNCOPUOS political meetings.

---

[1] Kuhn understood the scientific community as a group of individuals dealing with one particular discipline (Kuhn 1962: 180–182).

[2] Foucault regarded knowledge as a set of value, ideological and technical understanding (Foucault 1980: 137-139). Knowledge is an ability that makes people an expert and makes it possible to make accurate, effective and rationally justified decisions.

[3] By creating a new sub-theory, Adler and Haas responded to Keohan's 1989 remark that international relations lack the theoretical framework needed to explain the functioning of international politics under uncertainty (see below - Keohane 1989: 173). Through their theory, they tried to eliminate shortcomings and create a compromise between already existing positivist-empirical theories (neorealism, liberal institutionalism) and relativistic-cognitive approaches (cognitive analysis; see Haas 1992: 367-368; Adler - Haas 1992).

## THE CHARACTERISTICS OF THE EPISTEMIC COMMUNITIES AND THE SPECIFICATION OF THEIR PERFORMANCE IN INTERNATIONAL POLITICS

### Definition of the Epistemic Community

As already mentioned in the introduction to this text, Adler and Haas (Haas 1992; Adler - Haas 1992) began to deal with the importance of epistemic communities in international politics at the beginning of the nineties. Haas (1992: 3-5) defined the epistemic community as a network of professionals with a recognized level of expertise and competence in a certain thematic area, in which a set of normative, value and (2) causal beliefs are shared, (3) the concept of validity and (4) political conviction. The recognized level of expertise and competence allows this network of professionals to have an authoritative claim to assess the state of knowledge of the area. Epistemic communities, through their activities, illuminate the causes and consequences of complex international political problems, help states identify their interests and frame the themes of political negotiations (Haas 1992: 7-14). In other words, epistemic communities are networks of professionals who persuade other actors of international relations to take political decisions based on shared normative and causal beliefs within the epistemic community. The proposed form of political decisions must be strictly derived from expert knowledge, otherwise the epistemic community loses authority with other actors of international relations (Haas 1992: 16). The level of expertise of the epistemic community promotes shared criteria for assessing the correctness of the epistemic community and a common justification for political activity, which is usually an increase in human well-being. According to Haas (2010: 11579-11580), shared criteria for evaluating the epistemic community are important features that distinguish epistemic communities from other actors influencing political decisions (social movements, interest groups, etc. - Haas 1992: 17-19).

Haas's definition of epistemic communities in international relations has been criticized since its inception. Criticism, in the first place, pointed to the rethinking of the importance of epistemic communities in international politics. In their papers, Toke and Krebs pointed out that epistemic community research often mistakenly assumes the guaranteed access of epistemic communities to decision-making processes (Toke 1999: 97-102, Krebs 1999: 225-226). Furthermore, Krebs made a critical remark on "the better ability of epistemic communities to address complex political problems in comparison with governments" (Krebs 1999: 225). According to Toke, the epistemic community may not always offer a better solution to a complex political problem compared to the government, as unlike government officials are not in daily contact with the development of the problem (Toke 1999: 100-102). To date, the most serious critical remark is to question the very criteria (Dunlop 2000: 140-141), according to which Haas has defined the concept of epistemic communities. Dunlop notes that Haas attaches great importance to shared normative and causal beliefs, along with shared procedures for determining accuracy, but does not allow changes to these criteria. The Haas criteria also do not consider that members of international epistemic communities do not favour their personal or professional interests, the form of which is based on sub-norms and strategic interests of individual countries (Krebs 1999: 225,226). Cross (2013: 147-159) are trying to remove shortcomings of Haas's concept. Its refinement to the original model attaches an increased weight to the process of professionalization, which offers more detailed criteria for recognizing an organization associating epistemic communities and capturing its influence on other actors of international relations. The Cross' theoretical model is based on the assumption that the ability of the epistemic community to influence other actors of international relations depends mainly on the degree of internal coherence of the community, not on the high level of expertise or on the governmental (non-governmental) nature of an organization associating the epistemic community. Following the findings of social constructivism (Marsh - Rhodes 1992; Ruggie 1998; Wendt 1992; Schein 2010), emphasizing the importance of personal ties, shared values and the process of social interaction that determines the behaviour of the actors of international relations and their interests, socialization and links, and how to spread thoughts within the epistemic community. The inner coherence of a specific epistemic community determines a particular type of social interaction (so-called professionalization) that creates, enhances or renews the meaning and status of the profession (Cross 2013: 149). This process exists inside an organization associating epistemic communities, and within it infinitely creates and modifies common visions, shared standards, or professional identity itself.

According to Cross (2013), Epistemic communities are internationally associated within international governmental and non-governmental organizations. For the recognition of epistemic communities, it is important to follow in the governmental and non-governmental organizations the process of professionalization defined by three elements: (1) a common culture, (2) the selection of members and their further professional preparation, and (3) the quality and frequency of meetings. Based on these three elements, we will investigate in the next section whether we can designate professional organizations in a professional organization as epistemic communities.

*A Common Culture* Cross (2013: 150-151; 2011: 28-29), following Schein (2010: 14-16), defines the shared goal (the common values that the epistemic community seeks to achieve), symbolism (the common features the epistemic community characterizes it), the link (the shared heritage that the epistemic community passes on to the next generation) and shared identity (common mental models from which the members of the epistemic community are involved). The model of a common culture includes the "esprit de corps" (Cross 2011: 28), or between the epistemic community, there are friendly ties and devotion to the common goals of the epistemic community. Suppose there is a strong common culture within the epistemic community. In that case, members of the epistemic community are perceived as "one team" (Ibid: 29), so the probability of a high degree of community cohesion increases.

Professionalization within the pooling organization also strengthens the competitive selection of new members and the further personal development of already accepted members (Cross 2013: 150-151). Suppose the candidates for membership or members themselves also have intensive professional training. In that case, the epistemic community has a high degree of coherence and expertise. Finally, the third element of a professionalized epistemic community is the frequent and long-standing face-to-face encounter of individuals, which strengthens shared professional standards within the epistemic community, such as internal procedures, standards, the epistemic community protocol, and consensus building among members. Strengthening the links between the members of the epistemic community contributes mainly to informal (backstage) encounters in smaller groups where there is enough space for socializing and creating a common culture (Ibid: 150-151).

### Influence of epistemic communities in international politics

Demand for the activities of epistemic communities appears when political actors do not have the appropriate knowledge to assess the expected consequences of a political decision - that is, the problems of which the element of uncertainty is addressed (George 1980: 25–28; Adler - Haas 1992: 373, 375; Cross 2013: 151 –153). Uncertainty, whether perceived or real, is an integral part of international politics in the long run. It is visible in almost all areas of international politics, including political negotiations on global pandemics, mass migration, fighting pirates, or the way climate slows down. Hence, uncertainty has long been present in international politics and beyond, which we call "crisis" (Hay 1999: 317–335).

If uncertainty is present in the political negotiations, there is a demand for the activities of epistemic communities among actors (Haas 1992: 12-16). Thus, the epistemic communities gain the opportunity to influence international decision-making processes by bringing in new ideas and their further dissemination. A visible manifestation of the activities of epistemic communities in international politics is promoting the issue on the agenda of political action and later adopting a political decision in favour of epistemic communities (Adler - Haas 1992: 372-378).

In order to be able to challenge the political agenda and to persuade other actors of the need for a political decision in accordance with their interests, the epistemic communities must have access to the political decision-making process (ibid: 372-373). At the same time, they must convincingly present the problem they point out and want to resolve (Birklad 2011: 192; Kingdon 2014: 90–91). The approach of epistemic communities to the political decision-making process can be direct or indirect. The direct approach to the decision-making process in Alder and Haas (1992: 375-378) is in particular the involvement of members of epistemic communities in political decision-making and the transfer of responsibility for creating and applying policies to members of epistemic communities or to the epistemic communities themselves. The indirect approach of the epistemic communities to the decision-making process is to organize professional symposia during political discussions and to formulate opinions on political issues in the form of expert studies.

When it comes to how the problem is presented, the chances of the epistemic communities to raise the issue of political action and to persuade the actors of governance to apply a later political decision in favour of epistemic communities are increasingly used statistical indicators to present the problem. The effectiveness of this strategy is increasing when several indicators from different sources demonstrate the severity of the problem. The problem then becomes easier to attract political actors and a political decision is with the interest of epistemic communities. Examples of indicators may be changes in the value of unemployment, inflation, gross domestic product or birth rate and mortality (Birkland 2011: 192–193). In addition to the indicators themselves, how epistemic communities interpret the indicators is important to promote the problem to the agenda of political action and to later apply a political decision in favour of epistemic communities' interest in addressing a particular problem. Suppose political actors decide to collect data about a particular phenomenon. In that case, it means that they want to follow the evolution of the problem. Tracking then suggests that, from the perspective of these actors, the importance of the problem has increased (Kingdon 2014: 90-94).

The second element of the presentation, which significantly increases the chances of epistemic communities to raise the issue of political action and to convince political actors of the need to take a political decision in favour of epistemic communities, is "focusing events" (Birkland 2011: 180). Focusing events are "trigger mechanisms" that turn insignificant issues into issues that require major decisions. Signs of focusing events include their suddenness and relative scarcity, but the consequences of focusing events are large. Due to its characteristics, focusing events will gain rapid attention from political actors. Examples of focusing events are aircraft accidents, industrial accidents such as factory fires or oil spills (Ibid. 180-181). There are also important incentives for actors to pay attention to new issues or to pay more attention to existing but neglected topics. Suppose the epistemic communities use a focusing event to present the problem. In that case, it is more likely to remain on the agenda for political action and find a beneficial solution.

The hypothesis that epistemic communities' activity influences international space management development will be validated in two steps. In the first step, we will find out whether active actors in the space class political negotiations meet the criteria of the epistemic community (Haas 1992; Adler - Haas 1992; Cross 2013). In the second phase, we will verify whether the activity of epistemic communities influences the course of political negotiations. As criteria for their testing, we will use the criteria for introducing and disseminating new ideas into international politics (Haas - Adler 1992), which are the problem political action and the subsequent application of a political decision following the interest of the epistemic community. The hypothesis outlined will be entirely valid if (1) the actors involved in the space negotiations have the character of epistemic communities; (2) the problem of space drift has been a source of uncertainty that has prompted the demand for epistemic community activities; (3) where the space spill theme has been promoted to political action by representatives of epistemic communities; and (4) if a political decision has been made in accordance with the interest of epistemic communities.

## ACTORS INVOLVED IN POLITICAL ACTIONS ON SPACE DEBRIS
### Introduction to the problem of space debris and international efforts to solve it

Firstly, we will briefly present the problem of space debris and basic data on non-governmental and governmental organizations that deal with the international space debris problem. Space debriscollectively refers to all material that one has launched into outer space and is a non-functional cosmic object. Space debris is considered to be dysfunctional rockets, satellites and their parts, escaped fuel, or dropped objects of astronauts during ascension into space (Weeden 2011: 38–42). At low orbits, spacecraft move at high speeds (up to 8 km / s). If a low orbit space debris collides with a functional space device, then this collision can completely destroy it. Conversely, in a geostationary orbit in which the body orbits the Earth about once in about 24 hours, the apparent velocity of the space debris relative to the Earth's rotation is virtually zero. In a geostationary orbit, space debris tends to remain indefinitely, contributing to a faster depletion of a limited number of orbital positions on that path.

The United Nations Committee on the Peaceful Uses of Outer Space (UNCOPUOS) and the United Nations Office for Outer Space Affairs (UNOOSA) are primarily responsible for international space cooperation. UNCOPUOS is the UN General Assembly's political arena for peaceful cooperation between states in the use of outer space. Its activities take place in plenary and in two subcommittees - scientific and technical and legal. Both UNCOPUOS plenary and subcommittees decide on a consensus basis. UNOOSA[4] is responsible at the UN for the daily agenda of space exploration and exploitation. Its activities are managed by the Director and are divided into the Space Applications Section and the UNCOPUOS Services and Research Section. The Space Applications section is mainly provided by the United Nations Program on Space Applications (UNPSA), which promotes the use of space technologies in policy making. The UNCOPUOS Research and Services Section is involved in the preparation of internal research studies, which are subsequently submitted to UNCOPUOS meetings. In addition, this section is responsible for preparing UNCOPUOS negotiations and implementing UNCOPUOS decisions (UNOOSA 2014a). Thus, in order to promote the issue on the UNCOPUOS agenda for the benefit of epistemic communities, it is important for UNOOSA that their members gain positions in the UNCOPUOS Services and Research Section, where background documents for UNCOPUOS and its subcommittees are prepared, as well as (UNOOSA 2015d).

In particular, the following international non-governmental organizations are involved in the UNCOPUOS and UNOOSA spacecraft negotiations: International Astronautical Federation (IAF), Committee on Space Research (COSPAR),[5] International Institute of Space Law (International Institute of Space Law - IISL and Inter-Agency Space Debris Coordination Committee (IADC). While the IAF and COSPAR are dealing with the problem of space debris in Space (IAF 2014a: 5–6; COSPAR 2012a), IISL addresses the problem of space debris from the perspective of international space law. IADC is, unlike IAF, COSPAR and IISL, the International Intergovernmental Organization (IADC 2015), which space agencies contributing to the reduction of space debris. The following section answers the question whether these organizations meet the definition criteria of the epistemic community.

---

[4] UNOOSA is a specialized office of the UN Secretariat (UNOOSA 2014a).

[5] COSPAR is a scientific committee of the International Council for Science (ICSU), an international non-governmental organization bringing together national scientific institutions and international scientific unions (ISCU 2011: 2).

## International organizations as epistemic communities

### International Astronautical Federation

The non-governmental organization IAF aims to achieve a common goal of promoting space research, including the promotion of space technology development and deployment (IAF 2014a: 5-6). By supporting space research, the IAF makes a common reference by contributing to improve the life of society as a whole (ibid: 5). In addition to the unified title and acronym, the IAF considers its logo and trademark (IAF 2014b: 5) as elements of the IAF's common symbolism. In order to form a shared identity, the entry conditions for IAF membership candidates, according to which only a national organization having a common goal with the IAF, including space agencies, space industry, research or professional organizations, can apply for membership of the IAF.

The selection of IAF members is not competitive, but corresponds to professionalization (Cross 2013: 150–151). Instead of competing, the IAF decides to accept or not to accept a candidate in a multi-level admission process (IAF 2014a: 6-8). The applicant first submits the application to the IAF Office (ie the IAF's executive body), which will examine the application and subsequently pass it on to the IAF General Assembly. The IAF General Assembly then votes on acceptance or non-acceptance of the candidate by simple majority. The admission process emphasizes the assurance of coherence and expertise of the IAF, as each candidate for membership must - instead of previous training - provide the character of previous space research and exploitation activities. The qualification of a candidate for membership in the IAF is assessed at each stage of the multi-level admission process (IAF 2014a: 6-8). The IAF is competing for further personal development of already-accepted members, as the IAF annually grants its members several types of medals for significant achievements in space exploration and exploitation (IAF 2014c).

The International Astronautical Congress (IAC) is organized annually by the IAF to strengthen shared professional standards among its members. The Congress is the largest and most important international expert conference on space research and use (IAF 2014b). Usually IAC takes seven to eight days, during which there will be thirty symposiums, where individual IAF members meet. There is also opportunities for informal meetings during the IAC.

### Committee on Space Research

Like the IAF, COSPAR is working to achieve a shared goal of exchanging and collecting results, information and views on space research (COSPAR 2012a). By exchanging and collecting these results, COSPAR creates a common reference specified as generating scientific knowledge about outer space. The common symbolism of COSPAR is a single name, acronym and logo (Idem). To create a shared identity, a condition for membership candidates, according to which only national scientific institutions and the international scientific union in ISCU can apply for it.

As in the IAF, the selection of candidates in COSPAR is not competitive, but also here the selection of candidates is professional (Cross 2013: 105–151) as it is multi-level (COSPAR 2012b). First, a COSPAR candidate will submit an application to the COSPAR Council (ie the COSPAR Executive), extending the approved application to other COSPAR members. Suppose COSPAR members do not comment on the application within two months. In that case, the candidate will automatically become a full member of COSPAR (Idem). As in the IAF, the COSPAR multi-level recruitment process emphasizes COSPAR's coherence and expertise. A candidate for membership must submit a letter informing the COSPAR Council on space research activities so far in the represented territory, which then assesses the candidate's competence with COSPAR members (Idem).

COSPAR organizes several scientific conferences to strengthen shared professional standards among its members. The most important conference is the COSPAR Scientific Assembly (COSPAR 2014a), which takes place every two years and where members exchange information on the results of scientific research. In addition to the official program, numerous informal meetings take place at the COSPAR Scientific Assembly. In addition, COSPAR contributes to the strengthening of shared professional standards and standards by publishing the expert journal Advances in Space Research (COSPAR 2014b), which provides information on space research and thereby fulfills the COSPAR common goal.

### International Institute of Space Law

The activities of the NGO IISL seek to fulfil the common goal of developing the peaceful use of outer space (IISL 2013: 1). By fulfilling its goal, the IISL creates a common reference, which is international law in outer space. The IISL symbolism is a single name, acronym and emblem. The creation of a common identity is significantly supported by the admission process in which each candidate must first obtain a recommendation from three members or the IISL Director before filing an application (ibid: 2).

The selection of IISL members corresponds to professionalization because, as in the case of IAF and COSPAR, the IISL admission process is multi-level (ibid: 5-8; IISL 2014a: 2, 5). An applicant for membership first submits an application to the Committee for Membership of the IISL, who, after a simple majority, submits the application to the Board of Directors (IISL Executive Body). After approval of the application by a simple majority in the Board of Directors, the candidate for membership becomes a potential member of the IISL. Full membership can be requested by the IISL Committee no sooner than two years after becoming a potential member, while undergoing the same multilevel process as applying for potential member status. The admission process also contributes to ensuring the consistency and expertise of the IISL, as the candidate's professional profile is assessed at all stages. In addition to the admission process, another personal development of already accepted members is also competitive in IISL, as the IISL awards the Prof. Dr I.H.Ph. Diederiks-Verschoor (IISL 2014b).

To strengthen the shared professional standards of its members, the IISL organizes the Colloquium of International Space Law every year. The colloquium lasts three days, and the IISL members present the research results in international space law. Informal meetings are also taking place during the Colloquium (Idem). Shared professional standards are also reinforced by the publication of an occasional newsletter for IISL members.

### Inter-Agency Space Debris Coordination Committee

The IADC intergovernmental organization is pursuing a shared goal of exchanging and collecting information that contributes to reducing space debris (IADC 2011: 7-8, 17). The activity of the IADC also creates a common reference, which represents concrete technical procedures contributing to the reduction of space debris. The common IADC symbolism includes a single name, acronym, and logo. Shaped shared identities allow entry conditions under which only States, national or international space agencies that are active in space activities and are active in research to reduce the amount of space debris can be members of the IADC.

However, the selection of IADC members corresponds only partially to the concept of professionalization (Cross 2013: 150-151). There is no competitive or multi-level recruitment process at IADC (IADC 2011: 8, 17). The applicant's application is approved by IADC members by unanimous decision. However, IADC's consistency and expertise are emphasized in the recruitment process. When deciding to recruit a new candidate, the application is reviewed with information on his or her previous activities contributing to the reduction of space debris (ibid: 17). Only the State or Space Agency involved in the research contributing to the reduction of space debris may be a member of the IADC.

To strengthen the shared professional standards, the IADC annually organizes meetings of IADC working groups[6], which are always hosted by one of the members and is mainly attended by space debris experts (IADC 2014: 6). Furthermore, an IADC coordination meeting is held every year during the IAC. Communication between IADC members also takes place electronically through a closed communication system (IADC 2011: 7-8).

Overall, it is clear that NGOs and governmental associations of space and international law professionals meet the criteria of professionalization identified by Cross (2013), and are thus conducive to promoting their requirements in international arenas. This is summarized in Table 1.

---

[6] The particular IADC working groups are a working group on space debris measuring, for creating space debris layout models, for protecting against space debris, and for reducing the amount of space debris. Participation in the working group to reduce the amount of space debris is mandatory for each member of the IADC
the remaining working groups are voluntary (IADC 2011: 8-15).

**Table 1: Fulfilment of professionalization criteria by non-governmental organizations IAF, COSPAR, IISL and IADC**

| Criterion/Epistemic Community | Quality and frequency of meetings | Selection of members and their further professional preparation | Common Culture |
|---|---|---|---|
| IAF | Yes | Yes | Yes |
| COSPAR | Yes | Yes | Yes |
| IISL | Yes | Yes | Yes |
| IADC | Yes | Yes[7] | Yes |

Table 1 shows professionalisation process in all three international NGOs (IAF, COSPAR and IISL) and the IADC governmental organization (Cross 2013: 150-151). International NGOs are even professionalized in full. The non-governmental organizations IAF, COSPAR and IISL are thus organizations that associate epistemic communities.

In contrast, full-scale professionalization does not take place in the IADC government organization because the IADC does not decide on the selection of a candidate in a competitive or multi-level process. However, this deficiency is only partial since IADC assesses the competency of the candidate in the admission process. The IADC government organization is an organization associating epistemic communities. The partial diversion of the IADC recruitment process from the concept of professionalization can be reflected in the subsequent analysis by the IADC's reduced ability to influence other actors in space debris debates.

The difference between individual international organizations of epistemic communities is their kind. While IAF, COSPAR and IISL are NGOs, the IADC is an intergovernmental organization. In the UNOOSA and UNCOPUOS political space policy negotiations, only a partial consensus of the IADC epistemic community's standpoints with the epistemic community opinions emerging from the non-governmental organizations IAF, COSPAR, and IISL is expected to accentuate primarily professional interest. Indeed, the issue of solving the space debris problem is a controversial issue for government representatives in the IADC (Perek 2002).

## ACTIVITY OF EPISTEMIC COMMUNITIES IN POLITICAL NEGOTIATIONS ABOUT THE SPACE DEBRIS

The following section captures the influence of the activities of the epistemic communities IAF, COSPAR, IISL and IADC on political negotiations on space shifts in UNOOSA and UNCOPUOS from 1978 to the present. According to the Cross (2013: 151-152), the importance of epistemic communities is manifested in discussing issues that cause uncertainty about the consequences of a political decision. Adopting a political decision related to the use of outer space, including the problem of space debris, responds to these conditions. In the use of outer space, knowledge of several disciplines is necessary, and human's knowledge of the outer space is limited.

---

[7] Due to missing competitive procedure for selecting new members, selection new members and their further professional preparation correspond the professionalization partially only.

Epistemic communities are directly involved in the UNCOPUOS decision-making process, mainly in the framework of both subcommittees. Experts from the technical departments of government agencies, space agencies, national scientific institutions or universities are sent to the scientific and technical subcommittee within national delegations. Similar possibilities for direct involvement in the decision-making process have epistemic communities in the legal subcommittee, where countries within national delegations usually send experts from government departments, space agencies, national scientific institutions, or universities.

In both subcommittees, members of epistemic communities can submit their opinions and expert studies. Furthermore, the epistemic communities are involved in UNCOPUOS activities indirectly by obtaining consultative status within the Economic and Social Council (ECOSOC). This status allows NGOs and governmental organizations to participate in UNCOPUOS meetings and subcommittees in consultation. Besides that, NGOs and governmental organizations can enter into UNCOPUOS and sub-committee consultations indirectly on an ad hoc basis without ECOSOC consultative status, at the request of UNCOPUOS and sub-committees. Epistemic communities can indirectly engage in UNCOPUOS activities as well as by submitting studies and symposia during UNCOPUOS political talks and subcommittees (UNOOSA 2015c).

### Uncertainty and risks in the use of outer space

Members of the IAF Donald J. Kessler and Burton G. Cour-Palais (1978: 2640-2642) in the paper Collision Frequency of Artificial Satellites: The Creation of Debris Belt warned about the problem of space debris for the first time. They pointed out a growing number of satellites in orbits around the Earth, increasing the probability of further collisions between satellites. The functional and non-functional satellites, including their fragments, contributed to a higher probability of further collisions. An increasing number of collisions may increase the number of satellite fragments in the future, with each fragment likely to increase the likelihood of further collisions. A high number of collisions around the Earth could also cause the so-called caching of space debris, a situation where any collision creating a new amount of space debris will greatly increase the likelihood of a further collision.

Following the Kessler and Cour-Palais paper, representatives of epistemic communities have repeatedly pointed out the problem of space debris since the 1970s. UNCOPUOS and UNOOSA. The first to point out the uncertain consequences of the growing number of satellites around Earth for further space use (Kessler - Cour-Palais 1978) at the UN was the emeritus director of the Astronomical Institute of the Czechoslovak Academy of Sciences and then IAF and IISL member Luboš Perek, who held the post of Director of UNOOSA (2015b). In an internal UNOOSA study, Perek identified non-functional orbital devices and parts of them as a critical factor contributing to the likelihood of collision of functional orbital devices (United Nations 1979: 2-6).[8]

Another study referring to an article from 1978 is the COSPAR study (1981) on the effects of space activities on Earth and space. COSPAR presented the study during UNISPACE 1982, the UNOOSA and UNCOPUOS symposium, where the epistemic communities received indirect access to UNCOPUOS and UNOOSA. The study drew attention to the problem of space debris concerning the focusing event, the disintegration of the Soviet Cosmos 954 satellite and the subsequent fall of debris to unoccupied Canada in January 1978. The sudden crash of the Cosmos 954 caused large-scale damage, as the Cosmos 954 was powered by a nuclear reactor and fallen debris of satellite contaminated parts of Canada's territory with radioactive material (ibid: 17).

Thus, since the 1970s, the epistemic communities have pointed out to UNCOPUOS Scientific and Technical Subcommittee delegates the risks of increasing numbers of satellites around the Earth. The uncertainty that emerged then caused scientific and technological subcommittee demand for further activities of epistemic communities (expert studies) was manifested in 1987 when the Scientific and Technical Subcommittee delegates first requested the epistemic communities of the IAF and COSPAR to elaborate an expanded study with more detailed information on space debris (United Nations 1987: 30). Representatives of the IAF and COSPAR presented the desired study at the Scientific and Technical Subcommittee (United Nations 1989: 21-22), where they obtained direct access to UNCOPUOS.

---

[8] With its content on the interrelationships of space missions, the article follows the study of Kessler and Cour-Palais, which draws attention to the hypothetical state of the so-called space-cascading cascade (Kessler-Cour-Palais 1978: 2640-2642).

### Focusing events and use of statistical indicators

Graph 1 shows the evolution of the number of orbital objects recorded in orbits around Earth in 1957–2014 (NASA 2014: 10). The uppermost curve (I), on which the most critical shifts in the development of the debate on the space debris problem at UNOOSA and UNCOPUOS, and the focusing event (see below), are presented recorded objects in orbit around Earth. The lower curve (II) shows the total number of active space devices in orbits around the Earth. The difference between (I) the total number of objects in orbit around Earth and (II) the total number of active spacecrafts in orbits around the Earth indicates the total amount of space debris around the Earth.

**Graph 1: Progress on political negotiations about space debris in UNOOSA and UNCOPUOS and the amount of space debris around the Earth**

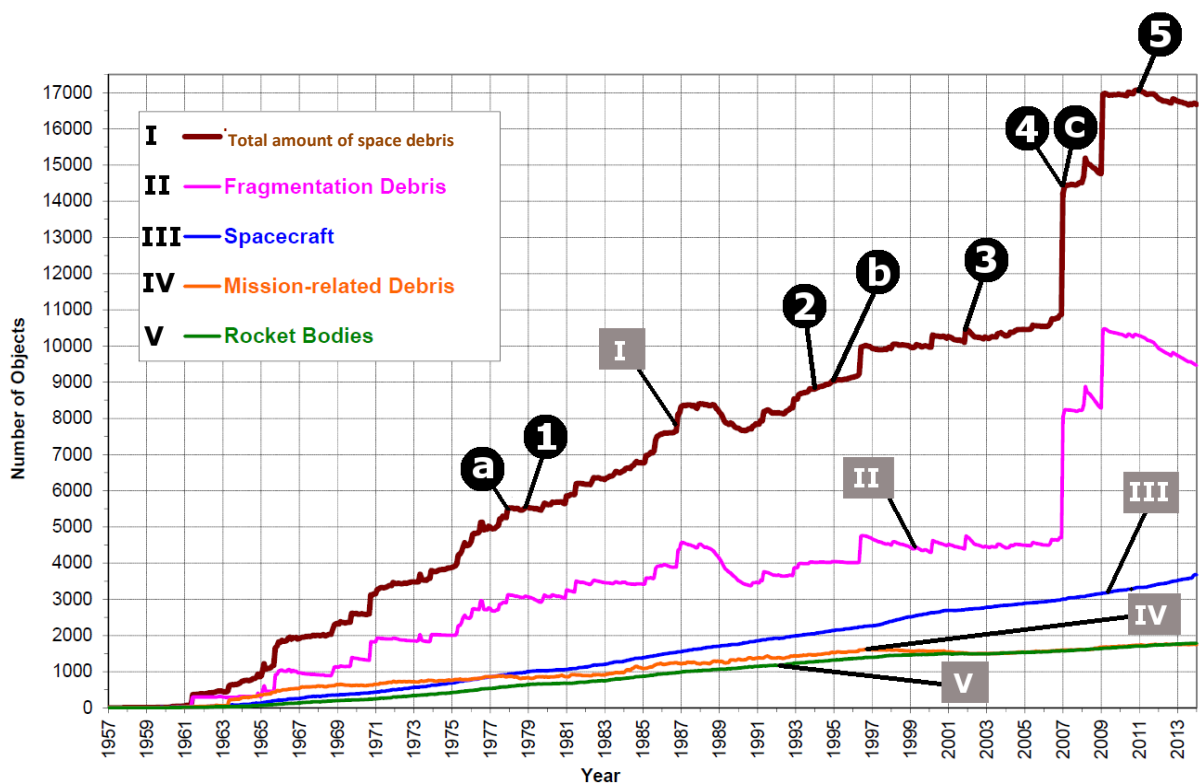

### Legend
1) **L. Perek** introduced study by OSAD 'Mutual Relations of Space Missions' in 1979
2) Space debris problem set on agenda in STS COPUOS
3) IADC took responsibility for elaborating technical solutions for space debris problem
4) UNGA adopted UN-Space Debris Mitigation Guidelines
5) Space debris problem set on agenda in LSC COPUOS

### Focusing events
a) Cosmos 954 accident
b) Cerise-Arianne collision
c) Anti-satellite missile test by PRC in 2007
d) Collision of Iridium 33 with space debris – non-functional satellite Cosmos 2251 (2009)

IAF and COSPAR have used space debris telescopic observations as statistical indicators to present the severity of the space debris problem. From their graphical presentation, it was evident that the amount of space debris in orbits around the Earth was exponentially growing (see Graph 1). The IAF and COSPAR have created a list of key focusing events for the presentation of the space problem. Focusing events have become the recorded breakage of satellites, along with data on the amount of newly created space spill. Studies have interpreted both the graphical processing of telescopic tracking values and the consequences of focusing events as indicators that the space debris problem is increasing (see Graph 1; COSPAR - IAF 1988: 6–11).[9]

Following the presentation of a joint IAF and COSPAR study, at least some of the UNCOPUOS Scientific and Technical Subcommittees have requested that the space debris issue be included on the agenda of the next subcommittee meeting, but other delegations have identified the issue as "preliminary" (United Nations 1989: 21-22 ). A joint study of the epistemic communities of the IAF and COSPAR of 1989 still did not create the necessary consensus among delegates of the Scientific and Technical Subcommittee on the need to discuss the issue of space debris at the next meeting. However, after its publication, at least some of the Scientific and Technical Subcommittee representatives have shown interest in further discussing the issue of space debris. At the same time as the IAF and COSPAR attempted to raise the issue of the Scientific and Technical Subcommittee, the IISL also sought to indirectly influence the decision-making process at UNCOPUOS by organizing a symposium on the environmental implications and responsibility of space use for members of the Subcommittee. During the IISL symposium, a contribution from the IAF member Nicholas Johnson on the direct dangers of space debris for further use of space has also appeared.

In addition to the numerical values indicating that the space debris problem is still growing, Johnson's study introduced space debris threats to specific orbits around the Earth. The study also showed that space debris is a complex problem (i.e., due to how the amount of space debris in the Earth's orbits is reduced), to which a comprehensive political decision will be needed (Johnson 1989: 482-486). Although the delegates concluded the subcommittee meeting, concluding that the issue of the space debris must be preceded by a consensus on a scientific and technical subcommittee, the presentation of Johnson's study was able to significantly increase the awareness of the delegates of the space subcommittee (Perek 2002: 127). Indeed, Johnson's debate has sparked a unexpected turbulent discussion among delegates of the legal subcommittee on the security aspects of space debris for the further use of space.

### Incorporating space debris into the UNCOPUOS agenda

Following a joint study by the IAF and COSPAR (1988), there was an interest in discussing space debris in UNCOPUOS and the UN General Assembly. In 1990 and 1991, it decided to include UNCOPUOS in the space debris theme for the next session (United Nations 1990: 93; United Nations 1991: 92). The UN General Assembly also urged that UNCOPUOS space debris negotiations cover all aspects related to the space debris problem that the epistemic communities have highlighted. Finally, after the application of the Scientific and Technical Subcommittee on the Integration of the Space Debris Problem into its Agenda (United Nations 1993a: 22), UNCOPUOS finally decided to address this issue at its next meeting (United Nations 1993b: 15, 1–23). Thus, in 1994, the problem of space debris first appeared among the permanent themes of the Scientific and Technical Subcommittee (United Nations 1994a: 12-13).

The situation in the legal subcommittee remained unchanged. However, because of the launch of space debris negotiations in the Scientific and Technical Subcommittee of the Legislative Subcommittees in 1994, they discussed the proposal to include the space debris problem at its meeting. However, the resulting debate led to the conclusion that the technical and scientific subcommittee must first evaluate all the technical aspects of the space debris (United Nations 1994b: 7-8).

---

[9] In absolute terms, the amount of space spill in orbits around the Earth has almost doubled in the period under review. While in the year 1980 there were about 4500 observed objects in orbits around the Earth, in 1987 this figure exceeded 7100, of which only 5% of the observed objects around the Earth were functional satellite equipment (COSPAR - IAF 1988: 6-10).

At a meeting of the UNCOPUOS Scientific and Technical Subcommittee in 1994, representatives of the IAF epistemic community presented a new study on the current state of space debris (IAF 1993: 1-24). It re-presented the severity of space debris problem using a statistical indicator that was the amount of space debris based on telescopic observations of objects in orbits around the Earth. From this, it was clear that the trend of the rapid increase of space debris against the situation during the 1980s significantly slowed down (see Graph 1). However, out of a total of 7,500 objects in orbit around Earth, 94% was space debris. Thus, the severity of the space debris problem remained high.

Following on from previous IAF, COSPAR and IISL studies, the IAF has also redefined the broad lines of technical solutions and specific technical solutions that will need to be taken into account in future political discussions at UNCOPUOS (IAF 1993: 3-9). IAF's specific technical solutions have been divided into two categories: technical solutions, on the one hand, to prevent the emergence of a further amount of space debris and, on the other hand, to eliminate the already existing amount of space debris (ibid: 1–24). By presenting the technical solutions of the IAF study, a political decision was made that, according to the epistemic communities IAF, COSPAR and IISL in UNCOPUOS, was to be adopted in with the solution to the problem of space debris. This political decision was intended to help prevent the emergence of another space debris and remove the already existing space debris.

The findings of the 1993 IAF study took into account the UNCOPUOS Scientific and Technical Subcommittee in 1995 in developing an effective plan for further space-related negotiations (United Nations 1995: 16). At the sessions of the Scientific and Technical Subcommittee in 1996-1998, the various aspects of the space debris problem identified by this study were discussed in details. Specifically, the Scientific and Technical Subcommittee from 1996 to 1998 dealt with the reliability of space debris measurement methods (United Nations 1996: 15–25), mathematical models of space debris spatial distribution (United Nations 1997: 19–32) and technical measures that reduce the amount of space debris (United Nations 1998: 7-9).

The scientific and technical subcommittee's discussions and decisions on the issue of space debris have become significantly more active in 1996–2000, when Dietrich Rex, Professor Technische Universität Braunschweig and a member of IAF and COSPAR, who has been dealing with the problem of space debris for a long time (United Nations 1996: 1-2). During the Rex Presidency, the Subcommittee on Science and Technology discussed a major focusing event - Cerise functional satellite collision (1995-033B) with a defective Ariane-1 missile (1986-019RF), i.e. the first clash of a cosmic device (United Nations 1997: 18). Information on this focusing event was prompted by the invitation of IADC epistemic community representatives to the S&T subcommittee (see below; Perek 2002: 129-130). Similarly, the collision of the Cerise with Ariane-1 as a focus event facilitated consensus on the text of the Final Space Report in Scientific and Technical Subcommittee in 1999.

The Final Technical Report (United Nations 1999: 4–29) reviewed the aspects of space debris discussed in studies conducted so far by the IAF and COSPAR epistemic communities and discussed by the subcommittees. It is clear from this that a consensus on a list of 16 reliable observation stations was created between 1996 and 1998 between the delegates of the Scientific and Technical Subcommittee, whose data UNCOPUOS will take into account in further discussions on space debris. Likewise, the Scientific and Technical Subcommittee delegates have reached a consensus over nine reliable models depicting the distribution of space debris around the Earth. The Final Space Report also referred to the basic outlines and starting points of concrete 12 technical solutions that will need to be taken into account when continuing the UNCOPUOS space debris debates.

### New international regulation of space debris problem

Although the epistemic communities from IAF, COSPAR and IISL have so far expressed the space debris problem in UNCOPUOS and UNOOSA, the responsibility for solving the space debris problem was suddenly transferred to IADC after the adoption of the final technical report by UNCOPUOS (United Nations 2002: 19-20). The transfer of responsibility for tackling the space debris problem at the IADC took place in 2002 when the IADC was granted an ad hoc consultative status for a Scientific and Technical Subcommittee meeting. The Scientific and Technical Subcommittee has instructed the IADC to develop technical solutions to prevent the emergence of space debris. Transferring responsibility to the epistemic community from the IADC has reformulated the space debris problem to the detriment of the demands of epistemic communities from non-governmental organizations IAF, COSPAR and IISL. This was reflected in the fact that the technical solutions commissioned by the UNCOPUOS commissioned the epistemic community of the IADC should only contribute to preventing the emergence of an additional amount of space debris, not to remove the already existing space debris (Ibid: 20).

The IADC Epistemic Community has requested a scientific and technical subcommittee to include only technical solutions to prevent the emergence of further space debris and soon confirmed it (IADC 2003: 7-16). The main reason why the proposed technical solutions did not address the problem of removing the actual amount of space debris, which has long been highlighted by non-governmental epistemic communities from IAF, COSPAR and IISL, was the anticipated disapproval of UNCOPUOS national delegations with other technical solutions due to their high cost and disruption security activities in outer space (Perek 2015).

Subsequently, the Scientific and Technical Subcommittee established a working group on Space debris to develop the UNCOPUOS directives for reducing space debris (hereafter "the Directive") on the basis of the proposed epistemic technical solutions community from IADC and comments from national delegations (United Nations 2004: 24). The working group on space debris existed between 2004 and 2006. The working group involved members of the epistemic communities from COSPAR and IAF Claudi Portelli and Petr Lála, who was a member of the Astronomical Institute of the Academy of Sciences of the Czech Republic. Lála and Portelli gradually held the position of chairman of the working group (United Nations 2004: 40, 2005a: 40, 2006: 39).

The working group drew up the Directives according to the technical solutions of the epistemic community of IADC and the comments of the national delegations UNCOPUOS, and the Scientific and Technical Subcommittee approved by the Directive (United Nations 2007a: 18-19). The Subcommittee approved them as a non-legally binding document (United Nations 2007b: 6-7). UNCOPUOS Plenum The directive was endorsed in 2007 (United Nations 2007c: 17) and definitively referred to the UN General Assembly as a non-binding resolution (United Nations 2007d: 4, 6). The Directives correspond to the technical solutions of the epistemic community of the IADC. By adopting the Directives by the UN General Assembly, the technical solutions to the epistemic community of IADC, which prevent the emergence of another space debris, from the technical solutions of the epistemic communities from IAF, COSPAR and IISL, which were to contribute to the elimination of the already existing space debris, were favored.

The UNCOPUOS and its subcommittees' space debris debates, which took place after the adoption of the UN General Assembly Directives, focused mainly on exchanging information on how to implement the National Directives. Since 2008, various technical solutions have been presented at scientific and technical subcommittee meetings to prevent the emergence of further space debris implemented by space agencies.

A UNOOSA report (United Nations 2008a: 5–6) was also presented at a meeting of the Scientific and Technical Subcommittee in 2008 confirming that a test of anti-satellite weapons of the People's Republic of China was held. It took place on January 2007 when the PRC rocketed its meteorological satellite with a medium-range missile. By shattering the meteorological satellite, 2500 fragments of space debris were created at low orbits around the Earth, implying an immediate 25% increase in the amount of space debris (see Figure 1). The sudden shoot-out of the Chinese meteorological satellite was - due to its unexpected and wide range of consequences - a major focal point for space policy political negotiations. Members of the epistemic communities of the IAF, IISL and IADC thus attempted to further reformulate the space debris problem at the meeting of the Scientific and Technical Subcommittee, taking into account the use of antisatellite weapons. However, there has been no reformulation of the space debris problem. UNCOPUOS has identified the use of anti-satellite weapons in its UNCOPUOS Plenary Subcommittee on Military Activities in Outer Space, which is on the agenda of the Conference on Disarmament,[10] not on the UNCOPUOS agenda and its subcommittees (United Nations 2008b: 3-7).

In recent years, the Scientific and Technical Subcommittee has continued to discuss the issue of space debris in collaboration with the epistemic community of IADC (United Nations 2009a: 14–15). It is only in those years that the further development of the space debris debate in the scientific and technical subcommittee seems to be most likely directed towards discussing technical solutions that will contribute to the removal of the already existing space debris. In 2013, IADC representatives introduced the latest space debris layouts models (IADC 2013a: 1–16). In the context of the 2009 focusing event, which was the sudden collision of the functional Iridium 33 satellite with the spacecraft - the dysfunctional satellite Cosmos 2251 - which increased the total amount of space debris in orbits by 20% (IADC 2013b: 5-8), the IADC recommended to the Scientific and Technical Subcommittee to discuss ways to remove large pieces of space debris from orbits around the Earth at further discussions. Following the reformation of the space debris problem, the first subcommittee's comments on the distribution of financial costs between countries were discussed at a meeting of the Scientific and Technical Subcommittee to remove large pieces of space debris from orbits around the Earth (United Nations 2013a: 15-16; United Nations) 2014a: 16).

The adoption of the Directives by the UN General Assembly and the subsequent General Assembly's call for the United Nations to continue its policy talks on space debris contributed to the inclusion of the space debris issue at the UNCOPUOS Legal Subcommittee 2009 (United Nations 2009b: 20-22). In the period 2009-2011, representatives of the epistemic communities of the IAF and the IADC delegates of the legal subcommittee presented the conclusions of the Scientific and Technical Subcommittee debate on the creation of the Directives (United Nations 2010: 20-22; United Nations 2011a: 19-21). A prolonged debate has been taking place in the Legal Subcommittee since 2012 on national legislation adopted by individual countries concerning space activities and its possible continuity with space debris (United Nations 2012: 21-24, 2013b: 20-23, 2014b: 22 –24).

\*\*\*\*

The text on the example of space negotiations at UNOOSA and UNCOPUOS has confirmed that the epistemic communities are influencing the evolution of the outer space administration. Firstly, international non-governmental organizations (IAF, COSPAR and IISL) and IADC government organizations involved in space debris negotiations consistently meet the criteria of professionalization (Cross 2013), and are therefore well placed to pursue their demands in international arenas. Secondly, the epistemic communities have long pointed out that the use of outer space brings risks and uncertainties associated with the problem of space debris. This uncertainty further strengthens the demand for activities of epistemic communities among UNCOPUOS scientific and technical subcommittee delegates. Thirdly, during the space debris debate, it is clear that without the activity of the epistemic communities associated with the NGOs IAF, COSPAR and IISL, this issue has probably never been on UNOOSA and UNCOPUOS. The activities of these communities have been on UNOOSA and UNCOPUOS since 1979 when epistemic communities have begun to work on the UNCOPUOS space debris agenda by providing expert studies and direct involvement in the decision-making process. It is also clear from the space debris debates that there was no partial lack of epistemic community from the IADC, an epistemic community, which, like the epistemic communities from international NGOs, was involved in the debate. Fourthly, the paper showed that the adoption of the Directives resulted in a partial political decision in line with the interest of the epistemic communities. The sudden transfer of responsibility for drafting the Directives to the IADC and the partial reformulation of the problem to the detriment of NGOs and purely professional organizations illustrate the lack of political representation of the represented states to take measures to reduce the amount of space debris. The adoption of such measures is associated with high financial costs and the possibility of disrupting military-security operations in outer space.

---

[10] The United Nations Conference on Disarmament has existed since 1979 as the UN Secretary-General's Office for Arms Control and Disarmament Agreements (UNOG 2014).

The limitation of the analysis presented here is that it cannot unambiguously prove the causal mechanism of the influence of epistemic communities on political space debris negotiations within UNCOPUOS and its subcommittees. By analyzing documents from the course of political debates alone, we cannot exclude other factors affecting the political negotiations on space debris (changing the UN agenda, the influence of other political actors in space agencies, advocacy groups, etc.). Nevertheless, the analysis of the problem of space debris suggests that more space should be devoted to the field of peaceful use of outer space in the research of epistemic communities (and more generally in the study of international relations). Further research should focus in particular on improving the success factors of the activity of epistemic communities in promoting their professionally grounded requirements in international arenas. The analysis presented here indicated that, in addition to professionalization, the institutional framework for the activities of epistemic communities, or the fact that it operates on a non-governmental or intergovernmental basis, is proven in this case. Different ways of participating in epistemic communities in particular decision-making processes also deserve special attention. This text showed different options for action before and during the agenda. He also marginally touched on the different positions from which epistemic community leaders can address the problem (experts; representatives of international organizations that represent arenas for political discussion of the problem; representatives of states in these arenas...), but due to the limited scope of these modalities follow in more detail.

**Note**

I want to thank my supervisor doc. Ing. Mgr. Štěpánka Zemanová, Ph.D., for valuable comments, advice and also for long-term support in writing the text. My thanks go to doc. Prof. RNDr. Luboš Perek, Dr.Sc., Dr. h. c. and Archive of the Academy of Sciences of the Czech Republic for comments and for making available archive materials from UNCOPUOS meetings. Last but not least, I would like to thank three anonymous opponents and the editorial office of the MV for their comments. Text loosely follows my diploma thesis, which was defended at the Departament of Politics and International Relations of the University of West Bohemia in Pilsen under supervision doc. Prof. PhDr. Šárka Cabadová Waisová, Ph.D.

Many thanks to Mgr. Václav Bartůšek for the English translation of the Czech version of paper.